\documentclass[10pt]{article}
\usepackage{euscript,amssymb,epsfig}
\newcommand{\be}[1]{\begin{equation}\label{#1}}
\newcommand{\ee}{\end{equation}}
\newcommand{\ba}[1]{\begin{eqnarray}\label{#1}}
\newcommand{\ea}{\end{eqnarray}}
\newcommand{\rf}[1]{(\ref{#1})}
\newcommand{\nn}{\nonumber}

\title{Massive scalar fields in the early Universe\footnote{Invited talk
presented at the 1st International Workshop on Astronomy and
Relativistic Astrophysics, 12-16 October, 2003, (IWARA2003),
Olinda-PE, Brazil; to appear in a special issue of the
International Journal of Modern Physics D.}}

\author{Uwe G\"unther$^a$\footnote{e-mail:
u.guenther@fz-rossendorf.de}~\footnote{present address: Research
Center Rossendorf, P.O. Box 510119, D-01314 Dresden, Germany} \
and Alexander Zhuk$^b$\footnote{e-mail:
zhuk@paco.net}~\footnote{address on leave: Department of Physics,
University of Odessa, 2 Dvoryanskaya St., Odessa 65100,
Ukraine}\\[2ex] $^a$ Gravitationsprojekt, Mathematische Physik
I,\\ Institut f\"ur Mathematik, Universit\"at Potsdam,\\ Am Neuen
Palais 10, PF 601553, D-14415 Potsdam, Germany
\\[1ex]
$^b$ Departamento de F$\acute{\mbox{\i}}$sica, Universidade
Federal de Para$\acute{\mbox{\i}}$ba\\ C. Postal 5008,
Jo$\tilde{\mbox{a}}$o Pessoa, PB, 58059-970, Brazil}

\date{04 November 2003}

\begin{document}
\maketitle

\begin{abstract}
We discuss the role of gravitational excitons/radions in different
cosmological scenarios. Gravitational excitons are massive moduli
fields which describe conformal excitations of the internal spaces
and which, due to their Planck-scale suppressed coupling to matter
fields, are WIMPs. It is demonstrated that, depending on the
concrete scenario, observational cosmological data set strong
restrictions on the allowed masses and initial oscillation
amplitudes of these particles.
\end{abstract}

PACS numbers: 04.50.+h, 11.25.Mj, 98.80.Jk

\vspace{.5cm}


\section{Introduction}

The concept of scalar fields is widely used in cosmology. Such
fields can have different origin, e.g. they can be Higgs fields in
4D GUT, dilaton fields in string theoretic  models, geometrical
moduli fields in different multidimensional setups. They can play
an important role in the early Universe because all of them, in
principle, can be considered as inflaton. On the other hand,  the
late time acceleration of the Universe can also be explained by
the presence of such fields (in quintessential models).
Additionally, they will be connected with various observable
phenomena so that cosmological and astrophysical data will
strongly constrain the parameters of these fields. In the present
paper, we shall mainly concentrate on gravitational excitons (for
short gravexcitons)/radions $\psi$ which are conformal
(geometrical moduli) excitations of the internal spaces in
multidimensional models \cite{GZ1}. We suppose that due to the
stabilization of the internal spaces, the gravexcitons acquire a
mass $m_{\psi}$. The characteristic feature of these particles is
their Planck scale suppressed coupling $1/M_{Pl}$ to standard
matter (SM) fields. As a result, the decay rate for their decay
into observable matter is small: $\Gamma \sim m_{\psi}^3/M_{Pl}^2
<< m_{\psi}$. From this point of view, gravexcitons are WIMPs
(Weakly-Interacting Massive Particles \cite{KT}). It is worth
noting, that a similar Planck scale suppressed interaction and
corresponding WIMP-like behavior holds for Polonyi particles in
spontaneously broken supergravity, scalarons in the $(R +R^2)$
fourth order theory of gravity  or moduli fields in the hidden
sector of SUSY. In the present work we demonstrate how
cosmological data can constrain gravexciton-related parameters for
different possible scenarios. In particular, we show how
restrictions can be derived on the masses of the gravexcitons as
well as on the amplitudes of their initial oscillations.

\section{Effective equation of motion}

The effective equation of motion for massive gravexcitons is (for
details see Ref. \cite{GSZ}):
\be{1.1} \ddot \psi + (3H+\Gamma)\dot \psi + m_\psi^2 \psi= 0\,
,\ee
where (by analogy with Ref. \cite{KLS1994}) we took into account
the effective decay of gravexcitons into ordinary matter, e.g.
into 4D photons:
\be{1.2} \Gamma \sim m_{\psi} \left(
\frac{m_{\psi}}{M_{Pl}}\right)^2 << m_{\psi}\, .\ee
In Refs. \cite{GSZ,kolb2} it was shown that the gravexciton
production due to interactions with matter fields is negligible
for the models under consideration. A corresponding source term on
the rhs of Eq. \rf{1.1} can, hence, be omitted.

The investigation of  Eq. \rf{1.1} is most conveniently started by
 a substitution
\be{1.3} \psi := B(t) u(t) := M_{Pl} e^{- \frac 12 \int
(3H+\Gamma)dt}u(t),  \ee
which, for constant $\Gamma$, leads to the following differential
 equation (DE) for the auxiliary function $u(t)$:
\be{1.4} \ddot u +\left[m^2_\psi-\frac 14 (3H+\Gamma)^2 - \frac 32
\dot H \right]u=0\, . \ee
This equation shows that at times when the Hubble parameter $H =
s/t \sim 1/t$ is less than the mass
\be{1.5}H < m_{\psi}\quad \Longrightarrow \quad t> t_{in} \sim
H^{-1}_{in} \sim \frac{1}{m_{\psi}}\, ,\ee
the scalar field oscillates
\be{1.6} \psi \approx C B(t) \cos (m_{\psi}t + \delta). \ee
The time $t_{in}$ roughly indicates the beginning of the
oscillations. Substituting the Hubble parameter $H = s/t$ into the
definition of $B(t)$ we obtain
\be{1.7}B(t) = M_{Pl}\; e^{- \frac 12 \Gamma t}\frac{1}{(M_{Pl}\;
t)^{3s/2}}\, ,\ee
where $s=1/2, 2/3$ for radiation dominated (RD) and matter
dominated (MD) stages, respectively. The parameter $C$ in Eq.
\rf{1.6} can be obtained from the amplitude of the initial
oscillation $\psi_{in}$:
\be{1.8} \psi_{in} \sim C B(t_{in}) \quad \Longrightarrow \quad
C_r \sim
\frac{\psi_{in}}{M_{Pl}}\left(\frac{M_{Pl}}{m_{\psi}}\right)^{3/4}\,
, \quad C_m \sim \frac{\psi_{in}}{m_{\psi}}.\ee
$C_r$ and $C_m$ correspond to particles which start to oscillate
during the RD and MD stages. Additionally, we took into account
that $\Gamma t_{in} \sim \Gamma/m_{\psi} << 1$.

Further useful differential relations are those for $B(t)$ in
\rf{1.3}, \rf{1.7} and for the energy density $\rho_{\psi}$ and
the number density $n_{\psi}$ of the gravexcitons. It can be
easily seen from the definition of $B(t)$ that this function
satisfies the DE
\be{1.9} \frac{d}{d t}\left(  a^3B^2 \right) = -\Gamma   a^3B^2 ,
 \ee
with $a(t) \sim t^{s}$ as  scale factor of the Friedmann Universe.
The energy density of the gravexciton field and the corresponding
number density, which can be approximated as
\be{1.10} \rho_{\psi} = \frac12 \dot \psi^2 + \frac12 m^2_{\psi}\,
\psi^2 \approx \frac12 C^2 B^2  m^2_{\psi}\, , \quad
n_{\psi}\approx \frac12 C^2 B^2  m_{\psi}, \ee
satisfy the DEs
\be{1.11} \frac{d}{d t}( a^3\rho _{\psi}) =-\Gamma a^3\rho _{\psi}
\qquad \mbox{\rm and} \qquad \frac{d}{d t}( a^3n_{\psi}) =-\Gamma
a^3n_{\psi} \ee
with solutions
\be{1.12}\rho _{\psi} \sim e^{-\Gamma  t} a^{-3}\qquad \mbox{and}
\qquad n_{\psi}\sim e^{-\Gamma  t} a^{-3}\, .\ee
This means that during the stage $ m_{\psi}> H \gg \Gamma $ the
gravexcitons perform damped oscillations and their energy density
behaves like a red-shifted dust-like perfect fluid $(\rho
_{\psi}\sim a^{-3})$ with slow decay $\sim e^{-\Gamma t}\sim 1$:
\be{1.13} \rho_{\psi} \approx \psi_{in}^2 m_{\psi}^2
\left(\frac{T}{T_{in}}\right)^3\, .\ee
$T_{in}$ denotes the temperature of the Universe when the
gravexcitons started to oscillate. According to the Friedmann
equation (the 00-component of the Einstein equation), the Hubble
parameter and the energy density (which defines the dynamics of
the Universe) are connected (for flat spatial sections) by the
relation
\be{1.14} H(t) M_{Pl} = \sqrt{\frac{8\pi}{3}\, \rho (t)} \sim
\sqrt{\rho (t)}\, .\ee
During the RD stage it holds $\rho (t)\sim T^4$ and, hence, $H^2
\sim T^4/M_{Pl}^2$. For gravexcitons which start their
oscillations during this stage, the temperature $T_{in}$ is now
easily estimated as
\be{1.15} H_{in} \sim m_{\psi} \sim \frac{T_{in}^2}{M_{Pl}} \quad
\Longrightarrow \quad T_{in} \sim \sqrt{m_{\psi}M_{Pl}}\, .\ee

If there is no broad parametric resonance ("preheating")
\cite{KLS1994}, then the decay plays the essential role when
$H\lesssim \Gamma$ and the evolution of the energy density of the
gravexcitons is dominated by an exponential decrease. The most
effective decay takes place at times
\be{1.16}t_D \sim H_D^{-1} \sim \Gamma^{-1} \sim
\left(\frac{M_{Pl}}{m_{\psi}}\right)^2m_{\psi}^{-1}\, .\ee

\section{Light and ultra-light gravexcitons: $m_{\psi} \leq 10^{-2}$GeV}

If the decay time $t_D$ of the gravexcitons exceeds  the age of
the Universe $t_{univ} \sim 10^{18}$sec then the decay can be
neglected. Eq. \rf{1.16} shows that this is the case for particles
with masses $m_{\psi} \leq 10^{21}M_{Pl} \sim 10^{-2}$GeV $\sim 20
m_e$ (where $m_e$ is the electron mass).

Subsequently, we split our analysis, considering separately
particles which start to oscillate before matter/radiation
equality $t_{eq} \sim H_{eq}^{-1}$  (i.e. during the RD stage) and
 after $t_{eq}$ (i.e. during the MD stage). According to
present WMAP data for the $\Lambda$CDM model it holds
\be{2.1} H_{eq} \equiv m_{eq} \sim 10^{-56}M_{Pl} \sim
10^{-28}\mbox{eV} \, .\ee

An obvious requirement is that gravexciton should not overclose
the observable Universe. This means that for particles with masses
$m_{\psi} > m_{eq}$ the energy density at the time $t_{eq}$ should
not exceed the critical density\footnote{It is clear that the
ratio between the energy densities of gravexciton and matter
becomes fixed after $t_{eq}$. If $\rho_{\psi}$ is less than the
critical density at this moment then it will remain undercritical
forever.}:
\be{2.2} \left. \sqrt{\rho_{\psi}}\; \right|_{t_{eq}\sim
H^{-1}_{eq}} \lesssim H_{eq} M_{Pl} \quad \Longrightarrow \quad
m_{\psi}\lesssim m_{eq}\left(\frac{M_{Pl}}{\psi_{in}}\right)^4\,
.\ee
Here we used the estimate $\rho_{\psi} \sim
(\psi_{in}/t)^2(m_{\psi}t)^{1/2}$ which follows from Eqs.
\rf{1.7}, \rf{1.8} and \rf{1.10}. For particles with masses
$m_{\psi} \gtrsim m_{eq}$,  relation \rf{2.2} implies the
additional consistency condition $\psi_{in} \lesssim M_{Pl}$, or
more exactly
\be{2.3} \psi_{in} \lesssim
\left(\frac{m_{eq}}{m_{\psi}}\right)^{1/4}M_{Pl} \lesssim M_{Pl}\,
.\ee

Let us now consider particles with masses $m_{\psi} \lesssim
m_{eq}$ which start to oscillate during the MD stage. From Eqs.
\rf{1.7}, \rf{1.8} and \rf{1.10} one finds for these particles
$\rho_{\psi} \sim (\psi_{in}/t)^2$ so that the inequality
\be{2.4} \left.\sqrt{\rho_{\psi}}\; \right|_{t_{in} \sim
H_{in}^{-1}} \lesssim H_{in}M_{Pl} \quad \Longrightarrow \quad
\psi_{in} \lesssim M_{Pl}\ee
ensures undercriticality of the energy density with respect to
overclosure of  the Universe.

It is worth noting that
the combination $-(9/4)H^2 - (3/2)\dot H $ in DE \rf{1.4} vanishes
for the MD stage because of $H = 2/(3t)$. Hence, for times $t \geq
t_{eq} \sim 1/m_{eq}$ the solutions of DE \rf{1.4} have an
oscillating behavior (provided that $m_{\psi} > \Gamma$) with a
period of oscillations $t_{osc} \sim 1/m_{\psi}$. For light
particles with masses $m_{\psi} \leq m_{eq}$ this implies that the
initial oscillations start at $t_{in} \sim 1/m_{\psi} $.

Finally, it should be noted that light gravexcitons can lead to
the appearance of a fifth force with characteristic length scale
$\lambda \sim 1/m_{\psi}$. Recent gravitational (Cavendish-type)
experiments (see e.g. \cite{experiment}) exclude fifth force
particles with masses $m_{\psi}\lesssim 1/(10^{-2}\mbox{\rm
cm})\sim 10^{-3}$eV. This sets an additional restriction on the
allowed mass region of gravexcitons. Furthermore, such
gravexcitons will lead to a temporal variability of the fine
structure constant above the experimentally established value
\cite{GSZ}. Thus, physically sensible models should allow for
parameter configurations which exclude such ultra-light
gravexcitons.


\section{Heavy gravexcitons: $m_{\psi} \geq 10^{-2}$GeV}
This section is devoted to the investigation of
gravexcitons/radions with masses $m_{\psi} \geq 10^{-2}$GeV for
which the decay plays an important role. Because of $m_{\psi}
>> m_{eq}$ the corresponding modes begin to oscillate during the RD stage.
We consider two scenarios separately. The first one contains an
evolutionary stage with  transient gravexciton dominance
($\psi$-dominance), whereas in the second one gravexcitons remain
always  sub-dominant.

\subsection{The transiently $\psi$-dominated Universe}
In this subsection we consider a scenario where the Universe is
already at the RD stage when the gravexcitons begin their
oscillations. The initial heating could be induced, e.g., by the
decay of some additional very massive (inflaton) scalar field. We
assume that the Hubble parameter at this stage is defined by the
energy density of the radiation. The gravexcitons start their
oscillations when the radiation cools down to the temperature
$T_{in} \sim \sqrt{m_{\psi}M_{Pl}}$ [see Eq. \rf{1.15}]. From the
dust-like red-shifting of the energy density $\rho_{\psi}$ [see
Eq. \rf{1.13}] follows that the ratio $\rho_{\psi}/\rho_{rad}$
increases like $ 1/T $ when $T$ decreases. At some critical
temperature $T_{crit}$ this ratio reaches $\sim 1$ and the
Universe becomes $\psi$-dominated:
\be{3.1} m_{\psi}^2 \psi_{in}^2 \left(\frac{T}{T_{in}}\right)^3
\sim T^4 \quad \Longrightarrow \quad T_{crit} \sim T_{in}
\left(\frac{\psi_{in}}{M_{Pl}}\right)^2\, .\ee
After that the Hubble parameter is defined by the energy density
of the gravexcitons: $H^2M_{Pl}^2 \sim \rho_{\psi}$ (with
$\rho_{\psi}$ from Eq. \rf{1.13}). This stage is transient and
ends when the gravexcitons decay at the temperature $T_D$:
\be{3.2} H_D^2 M_{Pl}^2 \sim \Gamma^2 M_{Pl}^2 \sim m_{\psi}^2
\psi_{in}^2 \left(\frac{T_D}{T_{in}}\right)^3\ \ \Longrightarrow \
\ T_D \sim T_{in}
\left(\frac{M_{Pl}}{\psi_{in}}\right)^{2/3}\left(\frac{m_{\psi}}{M_{Pl}}\right)^{4/3}\!
.
\ee
We assume that, due to the decay,  all the energy of the
gravexcitons is converted into radiation and that a reheating
occurs. The corresponding reheating temperature can be estimated
as:
\be{3.3} H_D^2 M_{Pl}^2 \sim \Gamma^2 M_{Pl}^2 \sim T_{RH}^4 \quad
\Longrightarrow \quad T_{RH} \sim
\sqrt{\frac{m_{\psi}^3}{M_{Pl}}}\, .
\ee
Because the Universe before the gravexciton decay was
gravexciton-dominated, it is clear that the reheating temperature
$T_{RH}$ should be higher than the decay temperature $T_D$. This
provides a lower bound for $\psi_{in}$:
\be{3.4} T_{RH} \geq T_D \quad \Longrightarrow \quad \psi_{in}
\geq \sqrt{m_{\psi}M_{Pl}} \sim T_{in}\, .\ee
Substitution of this estimate into Eq. \rf{3.1} shows that the
minimal critical temperature (at which the Universe becomes
$\psi$-dominated) is equal to the reheating temperature: $T_{crit
(min)} \sim T_{RH}$.

If we additionally assume the natural initial condition $\psi_{in}
\sim M_{Pl}$, then it holds $T_{crit} \sim T_{in}$ and the
Universe will be $\psi$-dominated from the very beginning of the
gravexciton oscillations. The upper bound on $\psi_{in}$ is set by
the exclusion of quantum gravity effects: $m_{\psi}^2\psi_{in}^2
\leq M_{Pl}^4$. Hence, in the considered scenario it should hold
\be{3.5} T_{in} \leq \psi_{in} \leq M_{Pl}
\left(\frac{M_{Pl}}{m_{\psi}}\right)\, .\ee

A successful nucleosynthesis requires a temperature $T \gtrsim
1$MeV during the RD stage. If we assume that this lower bound is
fulfilled for the reheating temperature \rf{3.3}, then we find the
lower bound on the gravexciton mass
\be{3.6} m_{\psi} \gtrsim 10^{4}\mbox{GeV}\, .\ee

It is also possible to consider a scenario where the $\psi-$field
acts as inflaton itself. In such a scenario, the Universe is
$\psi$-dominated from the very beginning and for the amplitude of
the initial oscillations one obtains:
\be{3.7} H M_{Pl} \sim \sqrt{\rho_{\psi}} \quad \Longrightarrow
\quad H_{in}M_{Pl} \sim m_{\psi}M_{Pl} \sim \psi_{in}m_{\psi}
\quad \Longrightarrow \quad \psi_{in} \sim M_{Pl}\, .\ee
The reheating temperature is then again given by the estimate
\rf{3.3} and the gravexciton masses should also fulfill the
requirement \rf{3.6}.

\subsection{Sub-dominant gravexcitons}

In this subsection we consider a scenario where the $\psi$-field
undergoes a decay, but the gravexcitons never dominate the
dynamics of the Universe. The Hubble parameter of the Universe is
then defined by the energy density of other (matter) fields which
behave as radiation for $t \leq t_{eq}$ and as dust for $t \geq
t_{eq}$. The energy density $\rho_{\psi}$ is always much less than
the total energy density of the other fields.

\subsubsection{Decay during the RD stage}
In this subsection, we analyze  the behavior of $\psi$-particles
that decay during the RD stage, when $H^2 \sim T^4/M_{Pl}^2$.
Again we will clarify for which masses $m_{\psi}$ and initial
oscillation amplitudes $\psi_{in}$ such a scenario can hold. A
decay during RD implies that the decay temperature $T_D$,
estimated as
\be{3.8} \Gamma \sim H_D \sim \frac{T_D^2}{M_{Pl}} \quad
\Longrightarrow \quad T_D \sim \sqrt{\frac{m_{\psi}^3}{M_{Pl}}}\,
,\ee
should be higher than the temperature
\be{3.9}T_{eq} \sim \sqrt{H_{eq}M_{Pl}} \sim 1\mbox{eV}\, \ee
of the matter/radiation equality. This yields the following
restriction on the gravexciton masses:
\be{3.10} T_D \gtrsim T_{eq} \quad \Longrightarrow \quad m_{\psi}
\gtrsim M_{Pl}\left(\frac{T_{eq}}{M_{Pl}}\right)^{2/3} \equiv m_d
\sim 1\mbox{GeV}\, .\ee
The mass parameter $m_d$ corresponds to particles which decay at
the moment $t_{eq}: \, \Gamma \sim H_{eq}$. The bound on the
initial oscillation amplitude $\psi_{in}$ can be found from the
energy sub-dominance condition for the gravexcitons at the moment
of their decay $t_D$
\be{3.11} \left. \rho_{\psi}\right|_{t_D} \approx \psi_{in}^2
m_{\psi}^2 \left(\frac{T_D}{T_{in}}\right)^3 < T^4_D\, . \ee
It reads
\be{3.12}  \psi_{in} < \sqrt{m_{\psi}M_{Pl}} \sim T_{in} \, ,\ee
where $T_{in}$ is defined by Eq. \rf{1.15}. It can be easily seen
that condition \rf{3.12} is supplementary to the condition
\rf{3.4}. During the decay, the energy of the gravexcitons is
converted into radiation: $\rho_{\psi} \rightarrow \rho_{r,2}$
with temperature $T_r^4 \sim \left. \rho_{\psi}\right|_{t_D}$. For
a scenario with $\left. \rho_{\psi}\right|_{t_D}<< T_D^4$, and
hence $T_r << T_D$, the energy density $\rho_{r,2}$ contributes
only negligibly to the total energy density and the gravexciton
decay does not spoil the standard picture of a hot Universe with
successful big bang nucleosynthesis (BBN).

\subsubsection{Gravexciton decay during the MD stage}

At the MD stage (for $t>t_{eq})$ the Hubble parameter reads (see
e.g. \cite{KT}, page 504)
\be{3.13} t \sim H^{-1} \sim
\frac{M_{Pl}}{T^{3/2}T_{eq}^{1/2}}\quad \Longrightarrow \quad H
M_{Pl} \sim T^{3/2}T_{eq}^{1/2}\, ,\ee
and the decay temperature $T_D$ of the gravexcitons can be
estimated as
\be{3.14} \Gamma \sim H_D \sim
\frac{T_D^{3/2}T_{eq}^{1/2}}{M_{Pl}}\quad \Longrightarrow \quad
T_D^{3} \sim \frac{(\Gamma M_{Pl})^2}{T_{eq}} \sim T_{eq}^3
\left(\frac{m_{\psi}}{M_{Pl}}\right)^4\left(\frac{m_{\psi}}{m_{eq}}\right)^2\,
.\ee
For a decay during MD this decay temperature should be less then
$T_{eq}$, and as implication a restriction on the mass of the
$\psi$-field can be obtained
\be{3.15} T_D < T_{eq} \quad \Longrightarrow \quad m_{\psi} <
M_{Pl}\left(\frac{T_{eq}}{M_{Pl}}\right)^{2/3} = m_d \, ,\ee
which is supplementary to the inequality \rf{3.10}. The
restriction on the initial amplitude $\psi_{in}$ can be found from
the condition of matter dominance and the fact that heavy
gravexcitons begin to oscillate at the RD stage when $T_{in} \sim
\sqrt{m_{\psi}M_{Pl}}$
\ba{3.16} && \psi_{in}^2m_{\psi}^2
\left(\frac{T_D}{T_{in}}\right)^{3} \sim T_r^4 < H_D^2M_{Pl}^2
\sim \frac{T_D^3T_{eq}}{M_{Pl}^2}M_{Pl}^2 \nn\\ && \Longrightarrow
\quad \psi_{in} < M_{Pl}\left(\frac{T_{eq}}{T_{in}}\right)^{1/2}
\sim M_{Pl}\left(\frac{H_{eq}}{m_{\psi}}\right)^{1/4} << M_{Pl}\,
.\ea
Condition \rf{3.16} guarantees that there is no additional
reheating and the BBN is not spoilt.

\section{Conclusion}
We discussed different cosmological scenarios affected by the
dynamics of gravitational excitons/radions. These massive moduli
fields describe the conformal excitations of the internal spaces
in higher dimensional models and are WIMPs in the external
spacetime. We demonstrated that observable cosmological data set
strong constraints on the gravexciton masses and the amplitudes of
their initial oscillations.

\section*{Acknowledgements}
AZ thanks C. Romero, V. Bezerra and the Department of Physics of
the Federal University of Para$\acute{\mbox{\i}}$ba for their kind
hospitality during his visit. He also thanks the Organizing
Committee of the workshop IWARA2003 for their kind invitation to
present a talk and for their financial support. Additionally, he
acknowledges financial support from CNPq during his stay in
Brazil.



\end{document}